\documentclass[12pt]{article}
\usepackage{amsmath,amsfonts,amsthm,amssymb}
\usepackage[top=3cm,bottom=3cm,left=3cm,right=3cm]{geometry}
\usepackage{graphicx}
\usepackage{lscape}
\newcommand{\CA}{\mathcal{A}}
\newcommand{\NR}{\mathbb{R}}
\newcommand{\p}{\partial}

\newcommand{\rref}[1]{(\ref{#1})}
\newcommand{\pd}[2]{\frac{\p #1}{\p #2}}

\newcommand{\bt}{\bar{t}}
\newcommand{\bx}{\bar{x}}
\newcommand{\ba}{\bar{a}}
\newcommand{\bu}{\bar{u}}
\newcommand{\ve}{\vec{e}}
\newcommand{\vm}{\vec{m}}
\newcommand{\vx}{\vec{x}}
\newcommand{\vu}{\vec{u}}

\newcommand{\rank}{\mathrm{rank}}
\newcommand{\disp}{\displaystyle}
\makeatletter
\renewcommand{\@biblabel}[1]{$^{#1}$}
\makeatother

\makeatletter
\long\def\unmarkedfootnote#1{{\long\def\@makefntext##1{##1}\footnotetext{#1}}}
\makeatother
\def \D {\hbox{d}}

\def\cs{\mathop{\rm cs}\nolimits}                                  
\def\ds{\mathop{\rm ds}\nolimits}                                  
\def\ns{\mathop{\rm ns}\nolimits}

\numberwithin{equation}{section}

\begin{document}

\begin{center}
\noindent {\bf \centering \Large Elliptic solutions of isentropic ideal compressible fluid flow
in (3+1) dimensions \\}
\end{center}
\vspace{5mm}
\begin{flushleft}
{\bf R. Conte$^{1,2}$, A.M. Grundland$^{3,4}$, B. Huard$^{5}$}\\

\vspace{5mm}
\noindent 1. LRC MESO, \'Ecole normale sup\'erieure de Cachan (CMLA) et CEA-DAM,
\\ 61, avenue du Pr\'esident Wilson, F-94235 Cachan Cedex, France.\\
2. Service de physique de l'\'etat condens\'e (URA 2464), CEA-Saclay, 
F-91191 Gif-sur-Yvette Cedex, France\\
3. Centre de Recherches Math{\'e}matiques, Universit{\'e} de Montr{\'e}al, \\
C.P. 6128, Succ.\ Centre-ville, Montr{\'e}al, (QC) H3C 3J7, Canada, \\
4. Universit{\'e} du Qu{\'e}bec, Trois-Rivi{\`e}res CP500 (QC) G9A 5H7, Canada \\
5. D{\'e}partement de math{\'e}matiques et de statistique, Universit\'e de Montr\'eal
C.P. 6128, Succ. Centre-ville, Montr{\'e}al, (QC) H3C 3J7, Canada \\
Email addresses : Robert.Conte@cea.fr, grundlan@crm.umontreal.ca, huard@dms.umontreal.ca
\end{flushleft}

\begin{abstract}
A modified version of the conditional symmetry method,
together with the classical method,
is used to obtain new classes of elliptic solutions of the isentropic
ideal compressible fluid flow in (3+1) dimensions.
We focus on those types of solutions which are expressed in terms of the
Weierstrass $\wp$-functions of Riemann invariants.
These solutions are of special interest since we show that they remain bounded
even when these invariants admit the gradient catastrophe.
We describe in detail a procedure for constructing such classes of solutions.
Finally, we present several examples of an application of our approach
which includes bumps, kinks and multi-wave solutions.  %BH Remplacé anti-bumps par kinks
\end{abstract}

% ===========================================================================

\section{Introduction}

The purpose of this paper is to construct bounded elliptic solutions of a compressible isentropic ideal flow in ($3+1$) dimensions.  Such solutions exist even in the case where the Riemann invariants admit the gradient catastrophe.

Let us first present a brief description of a procedure detailed in \cite{GH} for constructing rank-$k$ solutions in terms of Riemann invariants for the case of an isentropic compressible ideal fluid in $(3+1)$ dimensions.  Such a model is governed by the equations
\begin{equation}
\label{original-system}
 u^{\alpha}_t + \sum_{\beta=1}^4 \sum_{j=1}^3 {\CA^j}^{\alpha}_{\beta}(u) u^{\beta}_j = 0, \quad \alpha=1,2,3,4,
\end{equation}
where $\CA^1, \CA^2$ and $\CA^3$ are $4 \times 4$ real-valued matrix functions of the form
\begin{equation*}
\label{matrices}
\CA^j = \left(\begin{array}{cccc} u^i & \delta_{i1} \kappa^{-1} a & \delta_{i2} \kappa^{-1} a & \delta_{i3} \kappa^{-1} a \\ \delta_{i1} \kappa a & u^i & 0 & 0 \\ \delta_{i2} \kappa a & 0 & u^i & 0 \\ \delta_{i3} \kappa a & 0 & 0 & u^i
\end{array}\right), \quad j=1,2,3,
\end{equation*}
$\kappa = 2 (\gamma-1)^{-1}$ and $\gamma$ is the adiabatic exponent of the medium under consideration.  The independent and dependent variables are denoted by $x=(t=x^0,x^1,x^2,x^3) \in X \subset \mathbb{R}^4$ and $u=(a,\vec{u}) \in U \subset \mathbb{R}^4$, respectively, and $u_i$ stands for the first order partial derivatives of $u$, i.e. $u^{\alpha}_i \equiv \partial{u}^{\alpha} / \partial{x^i}$, $\alpha = 1,\ldots,4$, $i=0,1,2,3$.  Here, the quantity $a$ stands for the velocity of sound in the medium and $\vec{u}$ is the velocity vector field of the flow.  Throughout this paper, we adopt the summation convention over repeated lower and upper indices.

The purpose of this article is to obtain rank-$k$ solutions of system \rref{original-system} expressible in terms of Riemann invariants.  To this end, we seek solutions $u(x)$ of \rref{original-system} defined implicitly by the following set of relations between the variables $u^{\alpha}, r^A$ and $x^i$,
\begin{equation}
 \label{rank-k-solution}
\begin{split}
u = f(r^1(x,u), \ldots, r^k(x,u)), \quad r^A(x,u) = \lambda^A_i(u) x^i,\quad \ker{(\lambda_0^A \mathcal{I}_4 + \CA^i(u) \lambda^A_i)} \neq 0,
\end{split}
\end{equation}
for some function $f : \NR^k \to \NR^4$ and $A=1,\ldots, k \leq 3$. Such a solution is called a rank-$k$ solution if $\rank{(u^{\alpha}_i)} = k$.  The functions $r^A(x,u)$ are called the Riemann invariants associated with the wave vectors $\lambda^A = (\lambda^A_0,\vec{\lambda}^A) \in \NR^4$ of the system \rref{original-system}.  Here, $\vec{\lambda}^A = (\lambda^A_1,\lambda^A_2,\lambda^A_3)$ denotes a direction of wave
propagation and the eigenvalue $\lambda^A_0$ is a phase velocity of
the considered wave.  Two types of admissible wave vectors for the isentropic equations \rref{original-system} are obtained by solving 
the dispersion relation
\begin{equation}
\label{dispersion1}
\det{(\lambda_0(u) {\cal I}_4 + \lambda_i(u) \CA^i(u))} = 
[(\lambda_0 + \vec{u}\cdot\vec{\lambda})^2 - a^2\vec{\lambda}^2] 
(\lambda_0 + \vec{u}\cdot\vec{\lambda})^2                       
= 0.
\end{equation}
They are called the entropic $(E)$ and acoustic 
$(S)$ wave vectors and are defined by
\begin{equation}
\label{wavevectors}
\mathrm{i)}\,\, \lambda^E = (\varepsilon a+\vec{u}\cdot\vec{e},-\vec{e}), \quad \varepsilon = \pm 1,\quad
\mathrm{ii)}\,\, \lambda^S = (\det{(\vec{u},\vec{e},\vec{m})},-\vec{e} \times \vec{m}),
\quad |\vec{e}|^{\,2} = 1,
\end{equation}
where $\vec{e}$ and $\vec{m}$ are unit and arbitrary vectors, respectively.  

The construction of rank-$k$ solutions through the conditional symmetry method (CSM) is achieved by considering an overdetermined system, consisting of the original system \rref{original-system} in $4$ independent variables together with a set of compatible first order differential constraints (DCs), 

\begin{equation}
\label{Differential-constraints}
\xi_a^i(u) u^{\alpha}_i = 0, \quad \lambda^A_i \xi^i_a = 0, \quad a=1,\ldots,4-k,
\end{equation}
for which a symmetry criterion is automatically satisfied. Such notions as conditional symmetry, conditional symmetry algebra and conditionally invariant solution for the original system \rref{original-system} we use in accordance with definitions given in \cite{GH}. Under the above circumstances, the following result holds :

The isentropic compressible ideal fluid equations \rref{original-system} admit a $(4-k)$-dimensional conditional symmetry algebra $L$ if and only if there exists a set of $(4-k)$ linearly independent vector fields
\begin{equation}
\label{Vector-Fields-prop}
X_a = \xi^i_a(u) \pd{}{x^i}, \quad a=1,\ldots,4-k, \quad \ker{\left(\CA^i(u) \lambda^A_i\right)} \neq 0, \quad \lambda^A_i \xi^i_a = 0, \quad A=1,\ldots, k \leq 3,
\end{equation}
which satisfy on some neighborhood of $(x_0,u_0) \in X \times U$ the trace conditions
\begin{equation}
\label{trace-eq}
 \mathrm{i)}  \quad \mathrm{tr}{\left(\CA^{\mu} \pd{f}{r} \lambda\right)} = 0,  \quad
 \mathrm{ii)} \quad \mathrm{tr}{\left(\CA^{\mu} \pd{f}{r} \eta_{(a_1} \pd{f}{r} \ldots \eta_{a_s)} \pd{f}{r} \lambda\right)}=0, \quad \mu = 1,\ldots, 4,
\end{equation}
where
\begin{eqnarray*}
\label{threematrices}
& & \lambda = (\lambda^A_i) \in
\NR^{k \times 4},\quad r = (r^1,\ldots, r^k) \in
 \NR^k, \quad \frac{\p f}{\p r} = \left(
\frac{\p f^{\alpha}}{\p r^A}\right) \in \NR^{4 \times k}, \nonumber\\
\label{drdu}
& &\eta_{a_s} = \left(\pd{\lambda^A_{a_s}}{u^{\alpha}}\right) \in \NR^{k \times 4}, \quad s=1,\ldots,k-1,
\end{eqnarray*}
and $(a_1,\ldots,a_s)$ denotes the symmetrization over all indices in the bracket.  Solutions of the system which are invariant under the Lie algebra
$L$ are precisely rank-$k$ solutions of the form \rref{rank-k-solution}.

This result is a special case of the proposition in \cite{GH}.  Note that these symmetries
are not symmetries of the original system, but they can be used to construct solutions of the overdetermined system composed of \rref{original-system} and \rref{Differential-constraints}.

For the case of rank-$1$ entropic solution $E$, the wave vector $\lambda^E$ is a non-zero multiple of (\ref{wavevectors} i).  Therefore, the corresponding vector fields $X_i$ and Riemann invariant $r$ become
\begin{equation}
\begin{split}
&X_i = -(a + \vec{e} \cdot \vec{u})^{-1} e_i \frac{\p}{\p t} + \frac{\p}{\p x^i}, \quad i = 1,2,3, \\ 
&r(x,u) = (a + \vec{u} \cdot \vec{e}) t - \vec{e} \cdot \vec{x}, \quad |\vec{e}|^2 = 1,
\end{split}
\end{equation}
where we chose $\varepsilon = 1$ in (\ref{wavevectors} i).  Rank-1 solutions invariant under the vector fields $\{X_1,X_2,X_3\}$ are obtained through the change of coordinates
\begin{equation}
\label{changement-de-variable-rang-1}
\bar{t} = t,\, \bar{x}^1 = r(x,u),\, \bar{x}^2 = x^2,\, \bar{x}^3 = x^3,\, \bar{a} = a,\, \bar{u}^1=u^1,\, \bar{u}^2 = u^2,\, \bar{u}^3 = u^3,
\end{equation}
on $\mathbb{R}^4 \times \mathbb{R}^4$.  Assuming that the direction of the wave vector $\vec{e}$ is constant, the fluid dynamics equations \rref{original-system} transform into the system
\begin{equation}
\label{systeme-reduit-entropique}
\frac{\p \bar{a}}{\p \bar{x}^1} = \kappa^{-1} e_i \frac{\p \bar{u}^i}{\p \bar{x}^1}, \quad \frac{\p \bar{u}^i}{\p \bar{x}^1} = \kappa e_i \frac{\p \ba}{\p \bx^1}, \quad i=1,2,3,\\
\end{equation}
with the invariance conditions
\begin{equation}
\bar{a}_{\bar{t}} = \bar{a}_{\bar{x}^j} = 0, \quad \bar{u}^{\alpha}_{\bar{t}} = \bar{u}^{\alpha}_{\bar{x}^j} = 0, \quad j=2,3, \quad \alpha=1,2,3.
\end{equation}
The general rank-$1$ entropic $E$ solution takes the form
\begin{equation}
\label{rank-1-entropic}
\ba(\bt,\bx) = \ba(\bx^1), \quad \bu^i(\bt,\bx) = \kappa e_i \ba(\bx^1) + C_i,\quad C_i \in \mathbb{R},
\quad i=1,2,3,
\end{equation}
where the Riemann invariant $\bx^1 = r(x,u)$ is given by
$$r(x,u) = [(1 + \kappa)a + \vec{e} \cdot \vec{C}]t - \vec{e} \cdot \vec{x}, \quad \vec{C} = (C_1,C_2,C_3) \in \mathbb{R}^3.$$ 

A similar procedure can be applied to the rank-$1$ acoustic solution $S$.  Here, the wave vector $\lambda^S$ is a non-zero multiple of (\ref{wavevectors} ii) and the corresponding vector fields $X_i$ and Riemann invariant are
\begin{equation}
r(x,u) = \det{(\vec{u},\vec{e},\vec{m})} t - (\vec{e} \times \vec{m}) \cdot \vec{x}, \quad X_i = \frac{(\vec{e} \times \vec{m})_i}{\det{(\vec{u},\vec{e},\vec{m})}} \frac{\p}{\p t} + \frac{\p}{\p x^i}, \quad i=1,2,3.
\end{equation}
Again, the change of variables \rref{changement-de-variable-rang-1} leads to transformed dynamical equations, which we integrate in order to find rank-$1$ acoustic solution of the form
\begin{eqnarray}
&&\hspace{-2cm} \ba(\bt,\bx)=a_0, \quad \bu^1(\bt,\bx) = \bu^1(\bx^1),\quad \bu^2(\bt,\bx) = \bu^2(\bx^1), \quad C \in \mathbb{R},\\
&&\hspace{-2cm} \bu^3(\bt,\bx) = (e_1 m_2 - e_2 m_1)^{-1} \left[C - (e_2 m_3-e_3 m_2)\bu^1(\bx^1) - (e_3 m_1 - e_1 m_3)\bu^2(\bx^1)\right]. \nonumber
\end{eqnarray}
Here $\bu^1$ et $\bu^2$ are arbitrary functions of the Riemann invariant $\bx^1 = r(x,u)$ which has the explicit form
\begin{equation}
r(x,u) = C t - \det{(\vx, \ve, \vm)}.
\end{equation}

In general, the overdetermined system composed of (\ref{trace-eq} i) and (\ref{trace-eq} ii) is nonlinear and cannot always be solved in a closed form.
Nevertheless, particular rank-$k$ solutions for many physically interesting
systems of PDEs are well worth pursuing.
These particular solutions of (\ref{trace-eq} i) and (\ref{trace-eq} ii)
can be obtained by assuming that the function $f$
is in the form of a rational function, which may also be interpreted
as a truncated Laurent series in the variables $r^A$.  
This method can work only for equations having the Painlev\'e property
\cite{Conte-Painleve}.  Consequently, these equations can be very often integrated in terms of known functions.

Applying a version of the conditional symmetry method to the isentropic model (\ref{original-system}), several new classes of solutions have been constructed in a closed form \cite{GH1,GH}.  Comparing these results with the ones obtained via the generalized method of caracteristics (GMC) \cite{ZP}, it was shown that more diverse classes of solutions are involved in superpositions (i.e. rank-$k$ solutions) than in the case of the GMC \cite{GH}.

This paper is a continuation of the papers \cite{GH1,GH}.  The objective is to construct bounded elliptic solutions of the isentropic system (\ref{original-system}) using the version of the CSM proposed in \cite{GH}.  These types of solutions are obtained through a proper selection of differential constraints (DCs) compatible with the initial system of equations (\ref{original-system}).  That is, the solution should satisfy both the initial system \rref{original-system} and the differential constraints \rref{Differential-constraints}.  Among the new results obtained, we have rank-2 and rank-3 periodic bounded solutions expressed in terms of Weierstrass $\wp$-functions.  They represent bumps, kinks and multiple waves, all of which depend on Riemann invariants.  These solutions remain bounded even when the invariants admit a gradient catastrophe.  

%In addition, through a Monge-Amp\`ere type reduction, a rank-2 solution was found for an isentropic ideal fluid in which the sound velocity depends only on time.

The paper is organized as follows.  In Section 2 we construct rank-2 and rank-3 elliptic solutions of the system, among which multiple waves and doubly periodic solutions are included, and we show that they remain bounded everywhere.  Section 3 summarizes the results obtained and contains some suggestions for future developments.

% ===========================================================================

% ===========================================================================
\section{Rank-2 and rank-3 solutions}

The construction approach outlined in Section 1 has been applied to the
isentropic flow equations \rref{original-system}
in order to obtain rank-2 and rank-3 solutions.
The results of our analysis are summarized in Tables 1 and 2.
Several of them possess a certain amount of freedom. They depend on one or two arbitrary functions of one or two Riemann invariants,
depending on the case.
The range of the types of solutions obtained depends on different combinations
of the vector fields $X_a$ as given in \rref{Vector-Fields-prop}.
For convenience, we denote by $E_iE_j$,\, $E_iS_j$,\, $S_iS_j$,\, $E_iE_jE_k$, etc,
$i,j,k=1,2,3,$ the solutions which are the result of nonlinear superpositions
of rank-$1$ solutions associated with different types of wave vectors (\ref{wavevectors} i) and  (\ref{wavevectors} ii).
By $r^1,r^2$ and $r^3$ we denote the Riemann invariants which coincide
with the group invariants of the differential operators $X_a$ of the solution
under consideration.

The arbitrary functions appearing in the solutions listed in Tables 1 and 2
allow us to change the geometrical properties of the governed fluid flow
in such a way as to exclude the presence of singularities.
This fact is of special significance since, as is well known \cite{Mises,RJ},
in most cases, even for arbitrary smooth and sufficiently small initial data at $t=t_0$
the magnitude of the first derivatives of Riemann invariants
becomes unbounded in some finite time $T$.
Thus, solutions expressible in terms of Riemann invariants usually
admit a gradient catastrophe.
Nevertheless, we have been able to show that it is still possible in these cases
to construct bounded solutions expressed in terms of elliptic functions,
through the proper selection of the arbitrary functions appearing in the general solution.
For this purpose it is useful to select DCs corresponding to a certain class of the nonlinear Klein-Gordon equation which is known to possess rich families of bounded solutions \cite{Ab}. 
We choose elliptic solutions of the Klein-Gordon equation because a group theoretical analysis has already been performed \cite{WGT}.  The obtained results can be adapted to the isentropic ideal compressible fluid flow in ($3+1$) dimensions.  Thus, we specify the arbitrary function(s) appearing in the general solutions listed in Tables 1 and 2, say $\phi$, to the differential constraint in the form of the Klein-Gordon $\phi^6$-field equation in three independent variables $r^1,r^2$ and $r^3$ which form the coordinates of the Minkowski space $M(1,2)$
\begin{equation}
\label{Klein-Gordon}
 \phi_{r^1r^1} - \phi_{r^2r^2} - \phi_{r^3r^3} = c \phi^5, \quad c\in \mathbb{R}.
\end{equation}
Here, we choose $r^1$ to be timelike and $r^2, r^3$ to be spacelike coordinates.
It is well known (see e.g. \cite{WGT}) that equation \rref{Klein-Gordon} is invariant with respect
to the similitude Lie algebra $sim(1,2)$ involving the following generators
\begin{equation}
\begin{split}
\label{similitude-algebra}
&D = r^i \p_{r^i} - \frac{1}{2} \phi \p_{\phi}, \quad P_i = \p_{r^i}, \quad i=1,2,3,\\
&L_{ab} = r^a \p_{r^b} - r^b \p_{r^a}, \quad a \neq b = 2,3, \\
&K_{1a} = - (r^1 \p_{r^a} - r^a \p_{r^1}), \quad a=2,3,
\end{split}
\end{equation}
where $D$ denotes a dilation, $P_i$ represents translations, $L_{ab}$ stands for rotations and $K_{1a}$ for Lorentz boosts.
A systematic use of the subgroup structure \cite{WGT} of the invariance group
of \rref{Klein-Gordon} allows us to generate all symmetry variables $\xi$
in terms of the Riemann invariants $r^1,r^2,r^3$.
We concentrate here only on the case when symmetry variables are invariants
of the assumed subgroups $G_i$ of $Sim(1,2)$ having generic orbits of codimension one.
For illustration purposes, we perform a symmetry reduction analysis on four selected members of the list of subalgebras given in (\cite{WGT}, Table IV) which involve dilations.  For each selected subalgebra in Minkowski space $M(1,2)$, we compute the group invariants $\xi$ of the corresponding Lie subgroup and reduce the equation \rref{Klein-Gordon} to a second order ODE.  The application of the symmetry reduction method to equation \rref{Klein-Gordon}
leads to solutions of the form
\begin{equation}
\label{symmetry-reduction}
 \phi(r) = \alpha(r) F(\xi(r)), \quad r=(r^1,r^2,r^3),
\end{equation}
where the multiplier $\alpha(r)$ and the symmetry variable $\xi(r)$
are given explicitly by group theoretical considerations
and $F(\xi)$ satisfies an ODE obtained by substituting \rref{symmetry-reduction} into equation \rref{Klein-Gordon}.
The results of our computation are listed below.

\begin{equation}
\label{reduced-ODEs}
\begin{split}
& 1. \quad \{D, P_1\} : \quad \alpha = \{4c[(r^2)^2+(r^3)^2]\}^{-1/4}, \quad \xi = \frac{1}{2} \arctan{\frac{r^3}{r^2}}, \quad F'' + F + F^5 = 0,\\
& 2. \quad \{D, L_{31}\} : \quad \alpha = \{-c(r^1)^2/4\}^{-1/4}, \quad \xi = \frac{(r^2)^2+(r^3)^2}{(r^1)^2},\\
& \qquad \xi(1+\xi)F''+\left(2\xi+\frac{3}{2}\right)F' + \frac{3}{16} F + F^5 = 0, \\
& 3. \quad \{D + \frac{1+q}{q} K_{12}, L_{23}\} : \quad \alpha = \{-\frac{(2q+1)}{c}\}^{1/4}(r^1+r^2)^{q/2}, \\
& \qquad \xi = [(r^1)^2-(r^2)^2-(r^3)^2](r^1+r^2)^q, \\
& \qquad F''+\frac{3q+l}{2q+1} \frac{1}{\xi} F' + F^5 = 0,\quad  q=-l/3, l-2, 4-3l, \quad l\in \mathbb{Z}^+,\\
& 4. \quad \{D+\frac{1}{2}K_{12}, L_1 - K_{13}\} : \quad \alpha = (9/4C)^{1/4}\{r^2-(r^1+r^3)^2/4\}^{-1/2}, \\
& \qquad \xi = \frac{6(r^3-r^1)+6r^2(r^1+r^3)-(r^1+r^3)^3}{8(r^2-(r^1+r^3)^2/4)^{3/2}}, \quad (1+\xi^2)F''+\frac{7}{3}F'+\frac{1}{3}F + F^5 = 0. \raisetag{5cm}
\end{split}
\end{equation}
The parity invariance of \rref{Klein-Gordon} suggests the substitution
\begin{equation*}
F(\xi) = [H(\xi)]^{1/2}
\end{equation*}
which transforms the equations listed in \rref{reduced-ODEs} to
\begin{align}
\label{3.6p}
 &\{D, P_1\}&					&H'' = \frac{{H'}^2}{2H} - 2(H+H^3), \\
\label{3.7p}
 &\{D, L_{31}\}&					&H'' = \frac{{H'}^2}{2H} - \frac{1}{\xi(1+\xi)} \left[\left(2\xi+\frac{3}{2}\right)H' + \frac{3}{8}H + 2H^3\right],\\
\label{3.8p}
 &\{D + \frac{1+q}{q} K_{12}, L_{23}\}&		&H'' = \frac{{H'}^2}{2H} - \left[\frac{m}{\xi} H' + 2 H^3\right],
m=\frac{3 q + l}{2 q + 1}=(0,4/3,2),
\\
\label{3.9p}
 &\{D + \frac{1}{2} K_{12}, L_1 - K_{13}\}& 	&H'' = \frac{{H'}^2}{2H} - \frac{1}{1+\xi^2}\left[\frac{7}{3}\xi H' + \frac{2}{3}H + 2H^3\right],
\end{align}
where the three admissible values for the scalar $m$ come from group theoretical considerations \cite{WGT}.  Each of these four equations possesses a first integral
\begin{eqnarray}
& &
K'=\frac{1}{4} G g^2 \frac{{(g H)'}^2}{g H} - \frac{c_0}{4} (g H)^3 - 3 e_0 g H,
\label{eqfirstn}
\end{eqnarray}
in which the four sets of functions $G,g$ and constants $e_0,c_0$ obey
the respective conditions
\begin{eqnarray*}
& &
G=-\frac{3 c_0}{4},\ g^2= \frac{4 e_0}{c_0},\
\\
& &
G=-\frac{3 c_0}{4} \xi (\xi+1),\ g^2= -\frac{64 e_0}{c_0} \xi,\
\\
& &
G=-\frac{3 c_0}{4},\
(a,e_0,g^2)=
(0,0,k_1),\
(4/3,0,k_1 \xi^{4/3}),\
(2,e_0,-\frac{16 e_0}{c_0} \xi^2),\
\\
& &
G=-\frac{3 c_0}{4} (\xi^2+1),\
g=k_1(1+\xi^2)^{1/3},\
e_0=0,
\end{eqnarray*}
($k_1$ denotes an arbitrary nonzero real constant).
Under a transformation $(H,\xi) \to (U,\zeta)$
which preserves the Painlev\'e property,
\begin{eqnarray*}
& &
H(\xi) = U(\zeta) / g(\xi),\
\left(\frac{\D \zeta}{\D \xi}\right)^2=\frac{1}{G g^2},\
\end{eqnarray*}
the equation (\ref{eqfirstn}) becomes autonomous
\begin{eqnarray*}
& &
{U'}^2 - c_0 U^4 - 12 e_0 U^2 - 4 K' U=0,\ c_0 \not=0.
\label{eqfirsta} 
\end{eqnarray*}
When $K'=0$,
$U^{-1}$ is either a sine, cosine, hyperbolic sine or a hyperbolic cosine function,
depending on the signs of the constants,
therefore bounded solutions are easily characterized.

When $K'\not=0$,
it is convenient to first integrate this elliptic equation in terms of
the Weierstrass function $\wp(\zeta,g_2,g_3)$,
\begin{eqnarray*}
& &
U(\zeta)=\frac{K'}{\wp(\zeta) -e_0},\
g_2=12 e_0^2,\ g_3=-8 e_0^3 - c_0 {K'}^2, \nonumber\\
 & &\wp'^2 = 4(\wp-e_1)(\wp-e_2)(\wp-e_3) = 4\wp^3 - g_2\wp - g_3,
\end{eqnarray*}
(where we abbreviate $\wp(\zeta, g_2,g_3)$ by $\wp(\zeta)$) then to use the classical formulae which
connect $\wp$ and various bounded Jacobi functions.

This correspondence is quite easy to write down
if one uses the symmetric notation of Halphen \cite{Halphen} to represent the
Jacobi functions.
Halphen introduces three basis functions
\begin{equation*}
 h_{\alpha} (u) = \sqrt{\wp(u)-e_{\alpha}}, \quad \alpha=1,2,3,
\end{equation*}
and the connection between the Weierstrass $\wp$ function
and the Jacobi copolar trio $\cs,\ds,\ns$ is given by \cite[p.~46]{Halphen} 
\begin{equation*}
 \frac{\mathrm{cs}(z|k)}{h_1(u)} = \frac{\mathrm{ds}(z|k)}{h_2(u)}
 = \frac{\mathrm{ns}(z|k)}{h_3(u)} = \frac{u}{z} = \frac{1}{\sqrt{e_1-e_3}}, \quad k^2 = \frac{e_2 - e_3}{e_1 - e_3},
\end{equation*}
where $k$ is the modulus of the Jacobi elliptic functions.
For full details on Halphen's symmetric notation, see
\cite{MOS}.  We give here explicit solutions  
in terms of the Weierstrass function,
leaving the conversion to Jacobi's notation to the reader.
These solutions are obtained by convenient choices
of the normalization constants $e_0,c_0,K',k_1$.

With the normalization
$e_0=-1/3,\ c_0=-4/3,\ K'=C$,
the general solution of (\ref{3.6p}) is
\begin{equation}
\label{3.6sol}
F^2(\xi)= \frac{C}{\wp(\xi)+1/3},\
\zeta=\xi,\
g_2=\frac{4}{3},\
g_3=\frac{8}{27}+\frac{4}{3}C^2,\
C \in \mathbb{R}.
\end{equation}
With the normalization
$e_0=k_0^{-2}/48,\ c_0=-(4/3) k_0^{-2},\ K'=C$,
the solution of (\ref{3.7p}) has the form 
\begin{equation*}
F^2(\xi) = \frac{C \xi^{-1/2}}{\wp(\zeta) - \frac{1}{48 k_0^2}},\
\zeta=-2 k_0 \mathrm{argth} \sqrt{\xi+1},\
g_2=\frac{1}{192 k_0^4},\
g_3=-\frac{1}{13824 k_0^6}+\frac{4 C^2}{3 k_0^2},\
\end{equation*}
with $k_0, C \in \mathbb{R}$.

The three cases for equation \rref{3.8p} associated with the subalgebra $\{D + \frac{1+q}{q} K_{12}, L_{23}\}$ yield the respective solutions
\begin{equation*}
\begin{split}
& q = -k/3 :  
F^2(\xi) = \frac{C}{\wp(\xi)},\
\zeta = \xi,\
g_2=0,\
g_3=\frac{4 C^2}{3},
\\
& q = 4 - 3k : 
F^2(\xi) =\frac{C \xi^{-2/3}}{\wp(\zeta)},\
\zeta = 3 k_0 \xi^{1/3},\
g_2=0,\
g_3=\frac{4 C^2}{3 k_0^2},\\
& q=k-2 : 
F^2(\xi) =\frac{C \xi^{-1} }{\wp(\zeta)- \frac{1}{12k_0^2}},\
\zeta = k_0 \log {\xi},\
g_2=\frac{1}{12 k_0^4},\
g_3=-\frac{1}{216 k_0^6} + \frac{4 C^2}{3 k_0^2}.
\\
\end{split}
\end{equation*}

Finally, equation \rref{3.9p} integrates as (equation no $4$ in \rref{reduced-ODEs})
\begin{equation*} 
F^2(\xi) = \frac{C (\xi^2+1)^{-1/3}}{\wp(\zeta)},
\zeta = \xi\,\,\, {{}_2F_1}\left(\frac{1}{2},\frac{5}{6};\frac{3}{2};-\xi^2\right),\
g_2=0,\
g_3=\frac{4 C^2}{3k_0^2},
\end{equation*}
where ${{}_2F_1}$ denotes the hypergeometric function.

Using these results, we construct bounded rank-3 solutions of the equations
\rref{original-system}.
For this purpose, for each general solution appearing in Tables 2 and 3,
we introduce the arbitrary functions into the Klein-Gordon equation \rref{Klein-Gordon}
and select only the solutions expressed in terms of the Weierstrass $\wp$-function.

For illustration, let us now discuss the case of the rank-$3$ entropic solution $E_1E_2E_3$ which represents a superposition of three rank-$1$ entropic solutions $E_i$ given by \rref{rank-1-entropic}.  We assume that the entropic wave vectors $\lambda^{E_1}$, $\lambda^{E_2}$ and $\lambda^{E_3}$ are linearly independent and take the form
$$\lambda^{E_i} = (a + \vec{e}^{\,i} \cdot \vec{u}, - \vec{e}^i), \quad |\vec{e}^{\,i}|^2 = 1, \quad i=1,2,3.$$
Hence the corresponding vector fields $X_i$ and Riemann invariants are given by
\begin{equation}
X = \frac{\p}{\p x^3} - \frac{\det{(\vec{e}^{\,1}, \vec{e}^{\,2}, \vec{e}^{\,3})}}{\beta_3} \frac{\p}{\p t} + \frac{\beta_1}{\beta_3} \frac{\p}{\p x^1} + \frac{\beta_2}{\beta_3} \frac{\p}{\p x^2},\quad
r^i(x,u) = (a + \vec{e}^{\,i} \cdot \vec{u}) t - \vec{e}^{\,i} \cdot \vx,
\end{equation}
where $\beta_i = (\vec{e}^{\,2} \times \vec{e}^{\,3})_i (a + \vec{e}^{\,1} \cdot \vec{u}) + (\vec{e}^{\,1} \times \vec{e}^{\,3})_i (a+\vec{e}^{\,2} \cdot \vec{u}) + (\vec{e}^{\,1} \times \vec{e}^{\,2})_i (a+\vec{e}^{\,3} \cdot \vec{u})$.  The rank-$3$ entropic solutions invariant under the vector field $X$ are obtained through the change of coordinates
\begin{equation}
\bt = t, \bx^1 = r^1(x,u), \bx^2 = r^2(x,u), \bx^3 = r^3(x,u), \ba = a, \bu^1 = u^1, \bu^2 = u^2, \bu^3 = u^3,
\end{equation}
on $\mathbb{R}^4 \times \mathbb{R}^4$.  Specifying the form of the solution as a linear superposition of rank-$1$ solutions \rref{rank-1-entropic},
\begin{equation}
\ba = \ba_1(r^1) + \ba_2(r^2) + \ba_3(r^3), \quad \vec{u} = \kappa (\vec{e}^{\,1} \ba_1(r^1) + \vec{e}^{\,2} \ba_2(r^2) + \vec{e}^{\,3} \ba_3(r^3))
\end{equation}
the fluid dynamics equations \rref{original-system} transform to
\begin{equation}
\begin{split}
\sum_{i=1}^2 \sum_{j=i+1}^3 [\kappa (\vec{e}^{\,i} \cdot \vec{e}^{\,j})^2 + (1-\kappa) (\vec{e}^{\,i} \cdot \vec{e}^{\,j}) - 1 ]  \ba'_i(r^i) \ba'_j(r^j) = 0,
\end{split}
\end{equation}
while the invariance conditions have the form
\begin{equation}
\ba_{\bt} = \bu^1_{\bt} = \bu^2_{\bt} = \bu^3_{\bt} = 0.
\end{equation}
This solution exists if and only if the three entropic wave vectors $\vec{e}^{\,1}, \vec{e}^{\,2}, \vec{e}^{\,3}$ intersect at a certain specific angle given by
\begin{equation*}
 \cos{\phi_{ij}} = -\frac{1}{\kappa}, \quad i\neq j = 1,2,3,
\end{equation*}
where $\phi_{ij}$ denotes the angle between the wave vectors
$\vec{e}^i$ and $\vec{e}^j$ \cite{GH,ZP}.
Imposing the condition that each of the functions $\ba_i(r^i)$, $i=1,2,3$ obeys the ODE
$$F'' + F + F^5 = 0,$$
then according to \rref{3.6sol}, the rank-$3$ entropic solution takes the form
\begin{equation}
\label{isentropic-bounded}
\begin{split}
   &a = \sum_{i=1}^3
   \frac{C_i}{\left(\wp(r^i, \frac{4}{3}, \frac{8}{27} + \frac{4}{3} C_i^4)
   + \frac{1}{3}\right)^{1/2}}, \quad
 \vec{u} = \kappa \sum_{i=1}^3  \frac{C_i \vec{\lambda}^i}
 {\left(\wp(r^i, \frac{4}{3}, \frac{8}{27} + \frac{4}{3} C_i^4)
 + \frac{1}{3}\right)^{1/2}}, \\
  &r^i = -(1+\kappa) \frac{C_i}{\left(\wp(r^i, \frac{4}{3}, \frac{8}{27}
  + \frac{4}{3} C_i^4) + \frac{1}{3}\right)^{1/2}} t + \vec{\lambda}^i \cdot \vec{x},
  \quad i=1,2,3.
\end{split}
\end{equation}
Making use of an explicit expression for the zeros of the $\wp$-function, we show that the values for which the denominator in solution \rref{isentropic-bounded} vanish are not located on the real axis for a specific choice of the constants of integration $C_i$. Then we have the following result.

If the constants of integration $C_i$ are equal to $\sqrt{19}/{6}$, then the elliptic rank-$3$ entropic solution \rref{isentropic-bounded} of the isentropic ideal compressible fluid flow equations \rref{original-system} is bounded.

Indeed, according to recent results obtained by Duke and Imamoglu in \cite{Duke-Imamoglu}, the location of the zeros of the $\wp$-function can be given explicitly in terms of generalized hypergeometric functions.

Considering a lattice $\mathcal{L} = \mathbb{Z} + \tau \mathbb{Z}$, $\mathrm{Im} \, \tau > 0$, the doubly periodic Weierstrass $\wp$-function is defined by
\begin{equation*}
\label{wp-function}
\wp(z;\tau) = \frac{1}{z^2} + \sum_{\omega \neq 0} \left( \frac{1}{(z-\omega)^2} - \frac{1}{\omega^2}\right),
\end{equation*}
where the sum ranges over all $\omega \in \mathcal{L}$.  Note that the $\wp$-function assumes every value of the extended complex plane exactly twice in $\mathcal{L}$ and since it is even, its zeros are of the form $\pm z_0$.  The value of $z_0$ can be determined from the following theorem \cite{Duke-Imamoglu}.

{\it
\label{thm-wp-zeros}
The zeros $\pm z_0$ of the $\wp$-function are given by
\begin{equation}
\label{equation_for_z0}
z_0 = \frac{1+\tau}{2} + \frac{c_2 s^{1/4} {}_3 F_2 \left(\frac{1}{3},\frac{2}{3},1;\frac{3}{4}, \frac{5}{4};s\right)}{{}_2 F_1 \left(\frac{1}{12}, \frac{5}{12}; 1 ; 1-s\right)}, \quad c_2 = -\frac{i\sqrt{6}}{3\pi}, \quad |s|<1, \quad |1-s| < 1,
\end{equation}
where ${}_p F_q$ denotes the generalized hypergeometric functions and $s$ is a function of the modular discriminant $\Delta$ and the Eisenstein series $E_4$,
\begin{equation}
\label{intermediate_computations_z0}
s = 1 - \frac{1728 \Delta}{E_4^3}, \quad \Delta(\tau) = q \prod_{n\geq 1} (1-q^n)^{24}, \quad E_4(\tau) = 1 + 240 \sum_{n=1}^{\infty} \frac{n^3 q^n}{1-q^n} , \quad q = e^{2\pi i \tau}.
\end{equation}
It is understood that the principal branch is to be taken in the radical expression $s^{1/4}$. }

We now illustrate the application of this theorem by showing that the function $\wp(z,4/3, 1)$ is always strictly positive for real $z$.  This case is the specific case of solution \rref{3.6sol} for which $C = \sqrt{19}/6$.  From given invariants $g_2$ and $g_3$, by finding the roots $e_1, e_2, e_3$ of the cubic polynomial $4 t^3 - g_2 t - g_3$, one can determine the values of the periods $\omega_1$ and $\omega_2$.  In the case when $g_2 = 4/3$ and $g_3 = 1$, we obtain the periods $\omega_1 = 2.81$ and $\omega_2 = 1.405 + \mathrm{i} 2.902$, hence $\tau = \omega_2 / \omega_1 = 0.5 + \mathrm{i} 1.033$.

For any pair of periods $\omega_1, \omega_2$, the lattice $\omega_1 \mathbb{Z} + \omega_2 \mathbb{Z}$ can be rescaled in such a way that $\omega_1$ is normalized to $1$ by using the well-known formula for the $\wp$-function
\begin{equation}
\label{wp-rescaling}
\wp(z, \omega_1, \omega_2) = \wp(z/\omega_1, \omega_2/\omega_1) / \omega_1^2.
\end{equation}
Introducing the numeric value of $\tau$ into \rref{intermediate_computations_z0}, we can evaluate from \rref{equation_for_z0} and \rref{wp-rescaling} the zeros of the Weierstrass function $\wp(z,4/3,1)$, $z_0 = \pm (1.405 + \mathrm{i} 0.929),$ which are indeed complex.  Note that the $\wp$-function always possesses a double pole at $z=0$ and that it tends to $+\infty$ at this point.  Since it is continuous for all real $z \in (0,\omega_1)$, this implies that it must always be strictly positive.  Hence, $\wp(z,4/3,1) + 1/3 > 0$ on the real interval $(0,\omega_1)$ and the function $(\wp(z,4/3,1) + 1/3)^{-1}$ is therefore bounded for all real $z$.  Choosing the constants $C_i$ as values of the initial constant $C$ for which solution \rref{3.6sol} is bounded then guarantees the boundedness of functions $a$ and $\vec{u}$ in solution \rref{isentropic-bounded}.  Therefore, solution \rref{isentropic-bounded} is bounded everywhere, even when the Riemann invariants $r^i$ admit the gradient catastrophe.  A similar analysis can be applied to every solution presented in Table 3 to show that they are bounded.  This can be accomplished by the same procedure as presented above through the use of the theorem from \cite{Duke-Imamoglu}. 

\begin{figure}[h]
 \centering \includegraphics[width=12cm]{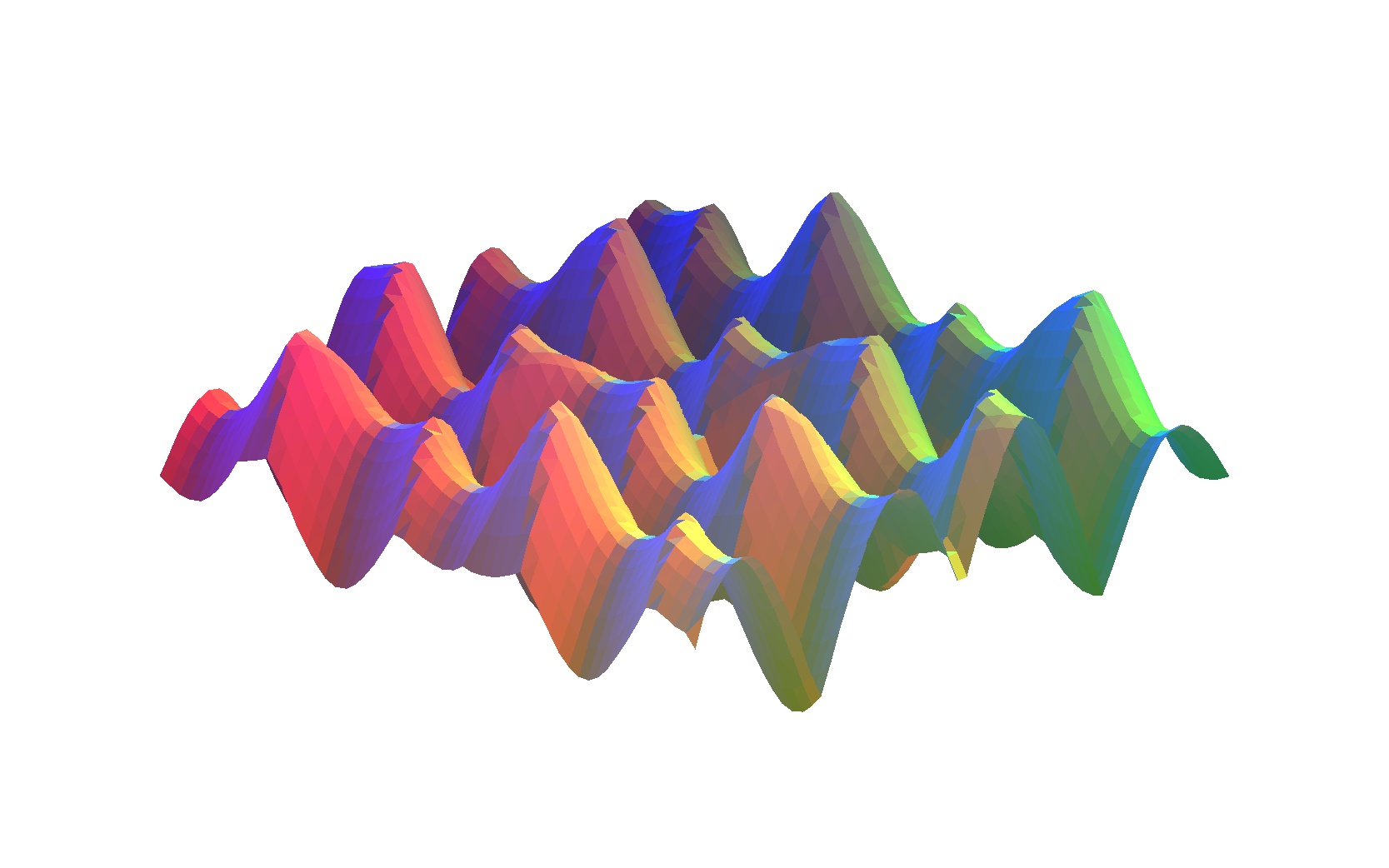}
\caption{Bounded entropic rank-$3$ solution}
\label{Rank-3-bounded-fig}
\end{figure}

This solution is physically interesting since it remains bounded
for every value of the Riemann invariants $r^i$.
Thus, it represents a bounded solution with periodic flow velocities (see figure \ref{Rank-3-bounded-fig}).
Similarly, it is possible to submit the arbitrary functions of
the differential equations no $2,3,4$ listed in \rref{reduced-ODEs} to obtain
other types of bounded solutions.
Table 3 presents these various types of solutions
with their corresponding Riemann invariants.
They are all bounded solutions of periodic, bump or kink type.
Note that these solutions of \rref{original-system} admit gradient
catastrophes at some finite time.
Hence, some discontinuities can occur like shock waves \cite{CF,F}.
Note also that the solutions remain bounded even when the first
derivatives of $r^i$ tend to infinity after some finite time $T$.
However, after time $T$, the solution cannot be represented in parametric
form by the Riemann invariants and ceases to exist.

% ===========================================================================
\section{Concluding remarks}

The methods presented in this paper can be applied quite broadly and can usually provide at least certain particular solutions of hydrodynamic type equations.  The conditional symmetries refer to the symmetries of the overdetermined system obtained by subjecting the original system \rref{original-system} to certain differential constraints defined by setting the characteristics of the vector fields $X_a$ to zero.  The conditional symmetries are not symmetries of the original system \rref{original-system}. However, they are used to construct classes of rank-3 solutions of this system which are not obtainable by the classical symmetry approach.  Among the new results obtained, we have rank-2 and rank-3 periodic solutions expressed in terms of the Weierstrass $\wp$-function that we have shown to be bounded over the real axis.  They represent bumps, kinks and multiple-wave solutions, all of which depend on Riemann invariants.  These solutions remain bounded even when the invariants admit the gradient catastrophe.  

%In addition, through a Monge-Amp\`ere type reduction, rank-2 solutions were found for an isentropic fluid flow in which the sound velocity $a$ depends only on time.  These solutions were obtained through the use of the conditional symmetry method and the Legendre transformation applied to the Monge-Amp\`ere equations.  

Among the questions that one may ask is what role do exact analytical solutions play in the physical interpretation.  One possible response is that such solutions may display qualitative behaviour which would otherwise be difficult to detect numerically or by approximations.  For example, the doubly periodic properties of certain solutions expressed in terms of the Weierstrass $\wp$-function would not be very easily seen numerically.

One could also inquire about the stability property of the obtained solutions.  Indeed, solutions which possess the property of stability should be observable physically and such analysis could be the starting point for perturbative computations. This task will be undertaken in a future work.\\

\noindent {\bf Acknowledgement}\\
The authors would like to thank William Duke (University of California, Los Angeles) for helpful and interesting discussion on the topic of this paper.  This work has been partially supported by research grants from NSERC of Canada. RC is pleased to thank the Laboratoire de physique math\'ematique du Centre de Recherches Math\'ematiques at the Universit\'e de Montr\'eal for its continuous support.

% ==========================================================================

%\bibliographystyle{plain}

\newpage
\begin{landscape}
\thispagestyle{empty}
{\footnotesize
\noindent {\bf Table 1. } Rank-2 solutions with the freedom of one, two or three arbitrary functions of one or two variables.  Unassigned unknown functions $a(\cdot ), u(\cdot ), \ldots$ are arbitrary functions of their respective arguments.

\begin{equation*}
\begin{array}{lllll}
\hline\hline
 No &\text{Type} & \text{Vector Fields} & \text{Riemann Invariants} & \text{Solutions}\\\hline

  1 & E_1 S_1 & X_1 = \frac{\p}{\p x^2} - \frac{\sigma_2}{\beta_1} \frac{\p}{\p t} - \frac{\beta_2}{\beta_1} \frac{\p}{\p x^1} & r^1 = ((1+k)\ba_1(r^1)+C_2)t - \ve^{\,1} \cdot \vx  &  \ba = \ba_1(r^1) +
 a_0, \quad [\vu_2, \ve^{\,2},\vm^2]=C \\
 & & X_2 = \frac{\p}{\p x^3} - \frac{\sigma_3}{\beta_1} \frac{\p}{\p t}
- \frac{\beta_3}{\beta_1} \frac{\p}{\p x^1} &r^2 = Ct - [\vx,\ve^{\,2},\vm^2], \quad [\ve^{\,1}, \ve^{\,2}, \vm^2] = 0 & \vu = k \ba_1(r^1) + \vu_2(r^2), \quad \bu^3_2 (r^2) = C_1 \bu^1_2(r^2)\\
& & \beta_i = -(\ve^{\,2} \times \vm^2)_i (a+\ve^{\,1}\cdot\vu) +
e_i^1[\vu,\ve^{\,2},\vm^2] & C_2 = (C_1e^1_1 - e_3^1)^{-1} & a_0, C, C_1, C_2 \in \mathbb{R}\\
& & \sigma_j = -e^1_1(\ve^{\,2} \times \vm^2)_j + e_j^1(\ve^{\,2} \times \vm^2)_1, j=2,3 & &\\
\hline

  2a & S_1 S_2 & X_1 = \frac{\p}{\p t} + u^1 \frac{\p}{\p x^1} +
u^2\frac{\p}{\p x^2} &r^1 = x^1 - u^1t & \ba=a_0, \hspace{5mm} \bu^1 = -\phi_{r^2},\hspace{5mm} \bu^2 = \phi_{r^1},  \\
 & & X_2 = \frac{\p}{\p x^3} & r^2 = x^2 - u^2t & \phi = \varphi(\alpha_1 r^1 + \alpha_2 r^2) + \beta_1 r^1 + \beta_2 r^2 + \gamma, \\
& & & & \bu^3 = \bu^3(r^1,r^2), \quad a_0, \alpha_i, \beta_i, \gamma \in \mathbb{R}, i=1,2,\\
 \hline

  2b & S_1 S_2 & X_1 = \frac{\p}{\p t} + u^1 \frac{\p}{\p x^1} + u^2 \frac{\p}{\p x^2} & 
r^1 = x^1-u^1t & \ba = a_0, \quad \bu^2 = \bu^3 = g(x^1-x^2), \quad a_0 \in \mathbb{R},\\
 & & X_2 = \frac{\p}{\p x^3} & r^2 = x^2-u^2t & \bu^1 = b(x^1 - tg(x^1-x^2),x^2-tg(x^1-x^2)) \\
\hline

  2c &  S_1 S_2 & X_2 = \frac{\p}{\p x^2} - \frac{\sigma_2}{\beta_1} \frac{\p}{\p t} - \frac{\beta_2}{\beta_1} \frac{\p}{\p x^1} & r^1 = \left(C_1 +
 \frac{\lambda^1_1}{\lambda^2_1}C_2\right)t - \vec{\lambda}^1 \cdot \vx &
 \ba = a_0, \quad a_0, C_1, C_2 \in \mathbb{R}\\
 & & X_3 = \frac{\p}{\p x^3} - \frac{\sigma_3}{\beta_1} \frac{\p}{\p t} - \frac{\beta_3}{\beta_1} \frac{\p}{\p x^1} & r^2 = \left(C_2 + \frac{\lambda^2_1}{\lambda^1_1}C_1 + G(r^1)\right)t
 - \vec{\lambda}^2 \cdot \vx & \bu^1 = \frac{1}{\lambda^1_1}(C_1 -
 \lambda^1_2\bu^2_1(r^1) - \lambda^1_3\bu^3_1(r^1)) \\ 
  & & \beta_j = \lambda^2_j[\vu,\ve^{\,1},\vm^1] - \lambda^1_j [\vu,\ve^{\,2},\vm^2] &\lambda^j_i = -(\ve^{\,j} \times \vm^j)_i & \qquad -
\left(\frac{\lambda^2_3}{\lambda^2_1} \eta + \frac{\lambda^2_2}{
\lambda^2_1}\right)\bu^2_2(r^2) + \frac{C_2}{\lambda^2_1}\\
 & & \sigma_i = \lambda^1_1 \lambda^2_i - \lambda^1_i \lambda^2_1 &G(r^1) = \frac{1}{\lambda^1_1} \left((\lambda^1_1\lambda^2_2 -
 \lambda^1_2\lambda^2_1)\bu^2_1(r^1)\right.  & \bu^2 =
 \bu^2_1(r^1)+\bu^2_2(r^2) \\
 & & & +
 \left. (\lambda^1_1\lambda^2_3-\lambda^2_1\lambda^1_3)\bu^3_1(r^1)\right) & \bu^3 = \bu^3_1(r^1) + \eta \bu^2_2(r^2), \quad \eta = \frac{\lambda^2_1\lambda^1_2 -
\lambda^1_1\lambda^2_2}{\lambda^1_1\lambda^2_3 -
\lambda^1_3\lambda^2_1}\\
\hline

  3 & E_1 E_2 S_1 & X = \frac{\p}{\p x^3} - \frac{\sigma_1}{\beta_{12}}\frac{\p}{\p t} +
\frac{\beta_{23}}{\beta_{12}} \frac{\p}{\p x^1} +
\frac{\beta_{31}}{\beta_{12}} \frac{\p}{\p x^2} & r^1 = \frac{\beta \bu^1_3(r^3)t - e^1_1 x^1 - e_2^1 x^2}{1 -
\alpha(1+\kappa)t} & \ba = \frac{\alpha((e^1_1 + e_1^2)x^1 + (e_2^1 + e^2_2)x^2)}{1-\alpha(1+\kappa)t}, \quad \bu^3 = u^3_0\\
 & & \sigma_1 = \varepsilon_{ijk} \, e_i^1e_j^2(\ve^{\,3}\times\vm)_k &  r^2 =  \frac{-\beta\bu^1_3(r^3)t - e_1^2 x^1 - e^2_2 x^2}{1 -
\alpha(1+\kappa)t} & \bu^1 = \frac{-\kappa\alpha \left(((e_1^1)^2 +
(e_1^2)^2)x^1 + (e^1_1e_2^1 + e_1^2e^2_2)x^2 \right)- \bu^1_3 (r^3)
}{1-\alpha(1+\kappa)t}\\
 & & \beta_{ij} = (e_j^1e_i^2 - e_i^1e_j^2)[\vu,\ve^{\,3},\vm^3]
      & r^3 = x^3 - u^3_0t & \bu^2 = \kappa\alpha\left(\frac{e_2^1\left(\beta
\bu^1_3(r^3)t -e^1_1x^1 -e^1_2x^2\right)}{1-\alpha(1+\kappa)t}\right. \\
 & & + (e_j^2(\ve^{\,3} \times \vm^3)_i -
    e_i^2(\ve^{\,3}\times\vm^{\,3})_j)(a+\ve^{\,1}\cdot\vu)  & \beta = (1+\kappa^{-1})/(e^1_1-e_1^2) & \hspace{0.8cm} \left. + \frac{ e^2_2\left(-\beta \bu^1_3(r^3)t -e_1^2x^1 -
e^2_2x^2\right)}{1-\alpha(1+\kappa)t}\right) + \frac{e^2_2 - e_2^1}{e_1^2
- e^1_1} \bu^1_3(r^3)\\
& & + (e_i^1(\ve^{\,3} \times \vm^3)_j -
    e_j^1(\ve^{\,3}\times\vm^3)_i)(a+\ve^{\,2}\cdot\vu) & & \alpha, u^3_0 \in \mathbb{R}\\

 \hline\hline
\end{array}
\end{equation*}
}
\newpage

\thispagestyle{empty}
{\footnotesize
\noindent {\bf Table 2. } Rank-3 solutions.  Unassigned unknown functions $a(\cdot ), u(\cdot ), \ldots$ are arbitrary functions of their respective arguments.

\begin{equation*}
\begin{array}{lllll}
\hline\hline
 \textit{No} & \text{Type} & \text{Vector Fields} &\text{Riemann Invariants} & \text{Solutions}\\\hline

1 &  E_1 E_2 E_3 & X_1 =\frac{\p}{\p x^3} + \frac{\sigma_1}{\beta_3} \frac{\p}{\p
    t} + \frac{\beta_1}{\beta_3} \frac{\p}{\p x^1} + \frac{\beta_2}{\beta_3} \frac{\p}{\p x^2} & r^i = (1+\kappa)a_i(r^i)t - \ve^{\,i} \cdot \vx, \,
i=1,2,3 & \ba =
\ba_1(r^1)+\ba_2(r^2)+\ba_3(r^3)\\
& &  \sigma_1 = -[\ve^{\,1},\ve^{\,2},\ve^{\,3}]   & \ve^{\,i} \cdot \ve^{\,j} = -1/\kappa, \, i\neq j = 1,2,3 & \vu =
\kappa(\ve^{\,1}
\ba_1(r^1) + \ve^{\,2} \ba_2(r^2)+ \ve^{\,3} \ba_3(r^3))\\
& & \beta_i = (\ve^{\,2} \times \ve^{\,3})_i(a+\ve^{\,1}\cdot\vu) & & \\
& & \hspace{7mm} + (\ve^{\,1}\times\ve^{\,3})_i(a+\ve^{\,2}\cdot\vu) & & \\
& & \hspace{7mm} + (\ve^{\,1}\times\ve^{\,2})_i(a+\ve^{\,3}\cdot\vu) & & \\
\hline

2a & E_1 S_1 S_2 &  X = e_1^2 \frac{\p}{\p x^1} + e^2_2 \frac{\p}{\p x^2}   & r^1 = ((1+k^{-1})f(r^1)+a_0+u^3_0)t-x^3 & \ba = k^{-1}f(r^1) + a_0, \quad \bu^1 = \sin{g(r^2,r^3)} \\
&  & & r^2 = t - x^1\sin{g(r^2,r^3)} + x^2\cos{g(r^2,r^3)} & \bu^2 =
 -\cos{g(r^2,r^3)}, \quad \bu^3 = f(r^1) + u^3_0\\
&  & &\frac{\p r^3}{\p t} + (f(r^1)+u^3_0)\frac{\p r^3}{\p x^3} = 0 & a_0, u^3_0 \in \mathbb{R} \\
\hline

2b & E_1 S_1 S_2 &  X = e_1^2 \frac{\p}{\p x^1} + e^2_2 \frac{\p}{\p x^2}   & r^1 = \frac{((1+k^{-1})B + a_0 + u^3_0)t - x^3}{1-(1+k^{-1})A t} & \ba = k^{-1}(A r^1 + B) + a_0,  \\
&  & & r^2 = t- x^1 \sin{g(r^2,r^3)} + x^2 \cos{g(r^2,r^3)} & \bu^1 = \sin{g(r^2,r^3)},  \bu^2 =
 -\cos{g(r^2,r^3)} \\
&  & & r^3 = \Psi\left[\frac{1}{A} (A(k a_0 - u^3_0 )t + x^3 -k a_0 - B) ((1+k)A t - k)^{-k/{k+1}}\right] & \bu^3 = A r^1 + B + u^3_0, \quad a_0, u^3_0 \in \mathbb{R}\\
\hline

2c & E_1 S_1 S_2 & X = \frac{\p}{\p x^3} & r^1 = (k^{-1}f(r^1)+a_0)t - x^1 \cos{f(r^1)} - x^2 \sin{f(r^1)} & \ba=k^{-1}f(r^1)+a_0, \quad \bu^1 = \sin{f(r^1)} \\
 &  & & r^2 = -t\cos{f(r^1)} - x^2 & \bu^2 = -\cos{f(r^1)}, \quad a_0 \in \mathbb{R} \\
 & & & r^3 = -t\sin{f(r^1)} + x^1 & \bu^3 =
 g(r^2\cos{f(r^1)} + r^3\sin{f(r^1))}\\\hline\hline
\end{array}
\end{equation*}
}
%\newpage
%\thispagestyle{empty}

\vspace{2cm}

\newpage
\thispagestyle{empty}

{\footnotesize
\noindent {\bf Table 3 : } Bounded real solutions for the nonscattering solution $E_1E_2E_3$ obtained by submitting the arbitrary functions to the various reductions \rref{3.6p}-\rref{3.9p} of the Klein-Gordon equation \rref{Klein-Gordon}.

\begin{tabular}{llll}
\hline\hline
 no & Riemann invariants & Solution & Type and comments\\\hline
 1 & $ \disp r^i = -(1+\kappa)\frac{C_i}{\left(\wp\left(r^i,\frac{4}{3},\frac{8}{27}+\frac{4}{3}C_i^4\right)+\frac{1}{3}\right)^{1/2}}t+\vec{\lambda}^i \cdot \vec{x}$
& $ \disp a = \sum_{i=1}^3 \frac{C_i}{\left(\wp\left(r^i,\frac{4}{3},\frac{8}{27}+\frac{4}{3}C_i^4\right)+\frac{1}{3}\right)^{1/2}}$ & Periodic solution \\
  & & $\disp \vec{u} = \kappa \sum_{i=1}^3 \frac{C_i \vec{\lambda}^i }{\left(\wp\left(r^i,\frac{4}{3},\frac{8}{27}+\frac{4}{3}C_i^4\right)+\frac{1}{3}\right)^{1/2}}$ & $C_i \in \mathbb{R}$ \\
\hline
% 2 & $\scriptstyle r^i = -(1+\kappa)\left((r^i)^{-1/2} \frac{C_i}{\wp\left(\zeta_i,12e_0^2,-8e_0^3 + 64  C_i^2 e_0\right) - e_0}\right)^{1/2}t+\vec%{\lambda}^i \cdot \vec{x}$ & $ \disp a = \sum_{i=1}^3 \left((r^i)^{-1/2} \frac{C_i}{\wp\left(\zeta_i,12e_0^2,-8e_0^3 + 64  C_i^2 e_0\right) - e_0}\right)^%{1/2}$ & \\
% & $\zeta_i = -2k_0\mathrm{arctanh}{\sqrt{r^i+1}}$ & $ \disp \vec{u} = \kappa \left(\sum_{i=1}^3 \left((r^i)^{-1/2} \frac{C_i}{\wp\left(\zeta_i,12e_0^2,-8e_0^3 + 64  C_i^2 e_0\right) - e_0}\right)^{1/2} \vec{\lambda}^i\right)$ & $e_0, \in \mathbb{R}, C_i >0$\\

2a & $r^i = -(1+\kappa)\left(\frac{C_i}{\wp\left(r^i,0,\frac{4 C_i^2}{3}\right)}\right)^{1/2}t + \vec{\lambda}^i \cdot \vec{x}$ & $ \disp a = \sum_{i=1}^3 \left(\frac{C_i}{\wp\left(r^i,0,\frac{4 C_i^2}{3}\right)}\right)^{1/2}$, $ \disp \vec{u} = \kappa \sum_{i=1}^3 \left(\frac{ C_i}{\wp\left(r^i,0,\frac{4  C_i^2}{3}\right)}\right)^{1/2}\vec{\lambda}^i$ & Periodic Solution \\
&  &  & $C_i >0$\\
\hline

2b & $r^i = -(1+\kappa)\left(\frac{C_i (r^i)^{-2/3}}{\wp(\zeta_i,0,\frac{4 C_i^2}{3 k_0^2})}\right)^{1/2} t + \vec{\lambda}^i \cdot \vec{x} $ & $\disp a = \sum_{i=1}^3 \left(\frac{C_i (r^i)^{-2/3}}{\wp(\zeta_i,0,\frac{4 C_i^2}{3 k_0^2})}\right)^{1/2}$, $ \disp \vec{u} = \kappa \sum_{i=1}^3 \left(\frac{C_i (r^i)^{-2/3}}{\wp(\zeta_i,0,\frac{4 C_i^2}{3 k_0^2})}\right)^{1/2} \vec{\lambda}^i$ &  Bump \\
& $\zeta_i = 3k_0 (r^i)^{1/3}$&  & $k_0 \in \mathbb{R}, C_i > 0$\\
\hline

2c & $\scriptstyle r^i = -(1+\kappa)\left(\frac{C_i (r^i)^{-1} }{\left(\wp\left(\zeta_i, 12 e_0^2, - 8 e_0^3 + 16 C_i^2 e_0 \right) - e_0\right)}\right)^{1/2}t + \vec{\lambda}^i \cdot \vec{x}$ & $\disp a=\sum_{i=1}^3 \left(\frac{C_i (r^i)^{-1} }{\left(\wp\left(\zeta_i, 12 e_0^2, - 8 e_0^3 + 16 C_i^2 e_0 \right) - e_0\right)}\right)^{1/2}$ & Bump\\
 & $\zeta_i = k_0 \ln{r^i}$ & $\disp \vec{u} = \sum_{i=1}^3 \kappa \left(\frac{C_i (r^i)^{-1} }{\left(\wp\left(\zeta_i, 12 e_0^2, - 8 e_0^3 + 16 C_i^2 e_0 \right) - e_0\right)}\right)^{1/2} \vec{\lambda}^i$ & $e_0,\in \mathbb{R}, C_i > 0$ \\
\hline

3 & $r^i = -(1+\kappa)\left(\frac{C_i ((r^i)^2+1)^{-1/3}}{\wp\left(\zeta_i,0,\frac{4C_i^2}{3k_0^2}\right)}\right)^{1/2}t + \vec{\lambda}^i\cdot\vec{x}$ & $\disp a = \sum_{i=1}^3 \left(\frac{C_i ((r^i)^2+1)^{-1/3}}{\wp\left(\zeta_i,0,\frac{4C_i^2}{3k_0^2}\right)}\right)^{1/2} $ & Kink\\
& $\zeta_i = r^i\,\,\, {{}_2F_1} \left(\frac{1}{2},\frac{5}{6};\frac{3}{2} ; -(r^i)^2\right)$ & $\disp \vec{u} = \kappa \sum_{i=1}^3 \left(\frac{C_i ((r^i)^2+1)^{-1/3}}{\wp\left(\zeta_i,0,\frac{4C_i^2}{3k_0^2}\right)}\right)^{1/2} \vec{\lambda}^i$ & $k_0,\in \mathbb{R}, C_i > 0$ \\\hline\hline
\end{tabular}
}
\end{landscape}

\end{document}